\def\year{2022}\relax
\documentclass[letterpaper]{article} 
\usepackage{aaai22}  
\usepackage{times}  
\usepackage{helvet}  
\usepackage{courier}  
\usepackage[hyphens]{url}  
\usepackage{graphicx} 
\usepackage{natbib}  
\usepackage{caption} 
\DeclareCaptionStyle{ruled}{labelfont=normalfont,labelsep=colon,strut=off} 
\usepackage{algorithm}
\usepackage{algorithmic}

\usepackage{newfloat}
\usepackage{listings}
\floatstyle{ruled}
\newfloat{listing}{tb}{lst}{}
\floatname{listing}{Listing}
\title{Improving Deep Localized Level Analysis: How Game Logs Can Help}
\author{
    Natalie Bombardieri, Matthew Guzdial
}
\affiliations{
    Department of Computing Science, Alberta Machine Intelligence Institute (Amii) \\
    University of Alberta, Edmonton, Alberta, Canada

    \{nbombard, guzdial\}@ualberta.ca
}

\usepackage{bibentry}

\begin{document}

\maketitle

\begin{abstract}
Player modelling is the field of study associated with understanding players. 
One pursuit in this field is affect prediction: the ability to predict how a game will make a player feel. 
We present novel improvements to affect prediction by using a deep convolutional neural network (CNN) to predict player experience trained on game event logs in tandem with localized level structure information. 
We test our approach on levels based on \textit{Super Mario Bros.} (\textit{Infinite Mario Bros.}) and \textit{Super Mario Bros.: The Lost Levels} (\textit{Gwario}), as well as original \textit{Super Mario Bros.} levels. 
We outperform prior work, and demonstrate the utility of training on player logs, even when lacking them at test time for cross-domain player modelling.
\end{abstract}


\section{Introduction}


Understanding how players react to video games helps game designers make better games \cite{harpstead2015drives}. 
The process of studying players to better understand them is called player modelling \cite{Yannakakis2013player}. Player modelling is a diverse field of study that ranges from prescriptive ``top-down" approaches where researchers propose a theory on how they think players ought to react, to interpretive, data-driven "bottom-up" approaches, where researchers try to find patterns in data, and then theorize on the significance of those patterns. 
The latter of these has, in our opinion, a greater chance of identifying more generally useful knowledge around player reactions.
However, many data-driven approaches still rely on hand-authored features, which requires expert design knowledge in terms of choosing which features to track.

Although the field of player modelling is diverse, in practice it is often used for limited purposes. A lack of available raw log information, caused in part due to ``game companies avoiding sharing game data with external researchers" \cite{lee2018game}, limits the number of works that make use of them. 
Prior work that analyzes raw game information includes \cite{liao2017}, who make use of unsupervised deep convolutional neural networks (CNNs) to analyze game event logs in concert with level structure data in order to predict self-reported player affect. 
Their system discovers interesting gameplay patterns that are associated with fun, frustration, and challenge, and they isolate and visualize specific player actions that are strongly associated with those emotions. 
However, we do not always have access to these player logs in all game domains, meaning this approach cannot be applied in many cases. 




In this paper, we introduce an approach that uses a CNN to predict self-reported player experience based on gameplay log data and localized level structure. We train our CNN on gameplay and level information from a \textit{Super Mario Bros.} clone, where we demonstrate higher levels of prediction accuracy using CNNs compared to prior work \cite{liao2017}. 
To evaluate the generality of the knowledge learned by our CNN we then test it on levels based on \textit{Super Mario Bros.: The Lost Levels}, as well as levels from the original \textit{Super Mario Bros}.
To the best of our knowledge, this represents the first instance of training on one game domain and testing on another, related domain without hand-authored transfer features or additional training.
We anticipate this is due to initially training the CNN on level chunks and unstructured game logs, which allows the model to learn relationships between log and level information, which it can then leverage even in distinct contexts without logs. 




\section{Related Work}

\subsection{Player Affect Modelling from Data}
A number of works aim to predict player affect by learning to identify patterns in game data.

Our work bears resemblance to \cite{liao2017}, who use the same \textit{Infinite Mario} dataset as we do, and use gameplay logs over a period of four time-steps. 
They also use level data, but feed in entire levels as input, rather than only small chunks as we do.
Our approach expands and improves on their approach by using the gameplay logs to calculate Mario's x-position, and only providing our CNN with a small window of the level information surrounding that position, which more closely relates to a player's experience of a subsection of a level as they play.

Other prior approaches have used gameplay logs from \textit{Infinite Mario} with level structure information to model player experience. \cite{pedersen2009modeling} summarize high-level information about level structure, such as enemy counts and number of gaps, whereas \cite{guzdial2016deep} inputs the level structure of whole levels into a CNN, similar to \cite{liao2017}. 
Both works track hand-authored features which summarize the log information of an entire level's play session, whereas our work uses raw game logs and subsets of level information from a few moments of gameplay.

We are not the first to attempt to transfer player modelling information between different games. 
The most common tactic to do this is to use summarized gameplay logs from dissimilar genres to predict player affect based on self reports \cite{shaker2015towards, melhart2021towards}. Specifically, they identify and track features shared between games as inputs for their machine learning models, and their models retain a high degree of accuracy even outside their training games. 
Approaches like these which generalize between games are rare, due to the requirement of having game logs for multiple games, as well as designer knowledge for both domains. 
Since our approach does not rely on human-authored features, it offers the possibility to both discover new relations between game logs that designers might not have otherwise considered, as well as removes the need for domain-specific expertise when analyzing games in this way.

We aim to solve the problem of lacking logs between different game domains by learning the relationship between log information and level information. 
An alternative to this could be addressed by meaningfully synthesizing game logs. This approach is used by \cite{luo2018player, luo2019making}, who make use of transfer learning to train CNNs to identify game events from gameplay video and output game logs, which can then be used for player affect prediction. Their approach could be used in tandem with ours to improve performance of models predicting player affect in other game domains. 


\cite{perez2011generic, yannakakis2010towards} make use of sensors to measure physiological responses to gameplay to model player experience. This removes the need for expert domain-specific designer knowledge, as the main feature being measured is the player's physical response to gameplay. A drawback to the approach is that it requires the use of intrusive sensors.
Our work instead uses gameplay logs and level structure information, which can be remotely collected without specialized hardware.


\subsection{Other Player Modelling Pursuits}
There are a number of other related approaches for extracting information on players. 
A common approach, demonstrated in \cite{drachen2009player, etheredge2013generic}, is to use summary game logs with clustering to discover player types, rather than to predict player experience. 
The knowledge identified from these clusters can be transferred to other games, but that is not its typical purpose. 

Experience-driven Procedural Content Generation (EDPCG) is a related area to our work \cite{yannakakis2011experience}, which describes approaches to adapt game content to a player in a live game setting. 
Historically, this involves some hand-authored adaptation approach \cite{yannakakis2009real}.
More recent work looks to automatically learn how best to adapt to different players \cite{shu2021experience}.
We view EDPCG as a related area, given that it also requires predicting player affect. 
However, EDPCG systems are typically applied to a single game, and not used across games. 

One of the most common alternatives to learning relationships of gameplay to human affect, is to hand-author a function to approximation human affect based on gameplay. 
This approach has seen use in content evaluation \cite{liapis2013towards,horn2014comparative}, tools \cite{liapis2013sentient}, and content generation \cite{tremblay2015algorithmic}.
This approach is limited by requiring extensive designer expertise in order to accurately capture relationships between player experience and gameplay.





\section{System Overview}

Our goal is to develop a system that can accurately predict player affect across similar levels, even for levels from different games. 
Our system makes use of a two-headed deep convolutional neural network (CNN) which uses gameplay logs as one head of the network and localized level structure information as the other, to learn to predict player self-reports of emotional response. 
This allows the CNN to learn relationship between these gameplay logs and localized level structure, which it can then leverage when it only has access to localized level structures for game levels outside its training data.

For training data we made use of the dataset provided by \cite{guzdial2016game}, which contains representations of 16 \textit{Infinite Mario} game levels \cite{togelius20102009}, one based on the original \textit{Super Mario Bros.} Level 1-1, and the other 15 generated by various PCG approaches. 
In addition, the dataset contains timestamped logs of game events from players playing those levels, player demographics, and self-reported player rankings of those levels. We provide our source code and datasets in a GitHub repository\footnote{https://github.com/gnatbomb/IDLLA}.

\begin{figure*}
    \centering
    \includegraphics[width=\textwidth, keepaspectratio]{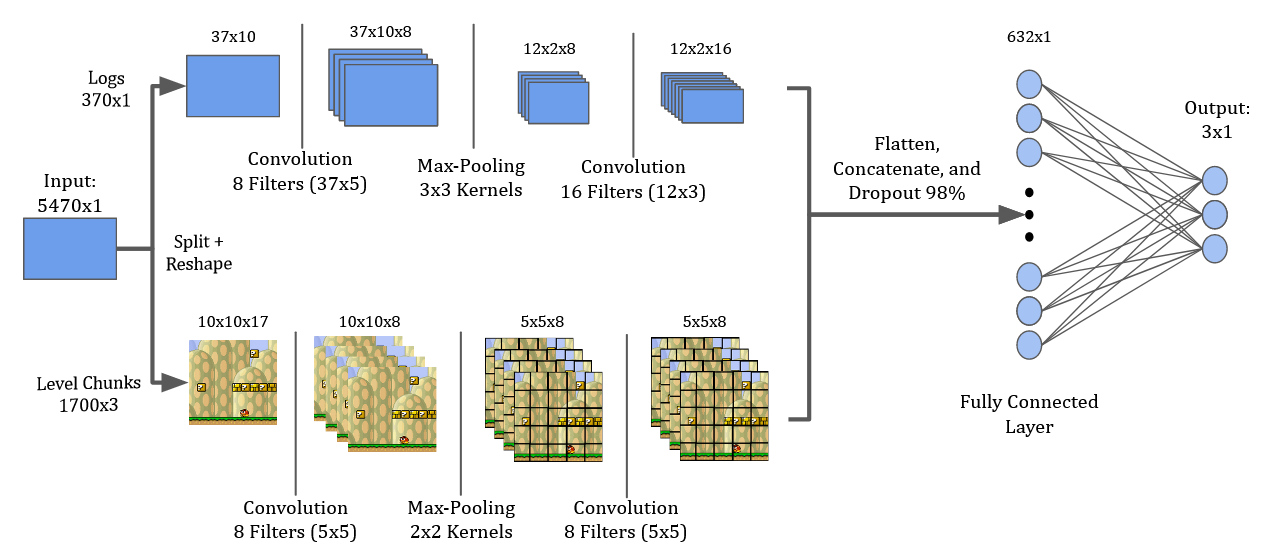}
    \caption{\small Visualization of our two-headed neural network architecture. Log data (37 values over 10 time-steps) and level chunks (three 10x10x17 chunks) are fed in together before being split into our model's  two heads. The resulting data are fed back into a single fully-connected layer of size 632x1, then again into a 3x1 fully-connected output layer. The level screenshot visualizations along the bottom of this figure are intended as an approximation of the level structure information for visual clarity, and do not represent the literal formatting of this data in our model.\normalsize }
    \label{fig:model_visualization}
\end{figure*}
\subsection{Data Parsing - Logs}

We re-represent the given gameplay logs and demographic information as an array. 
Each array represents a single time-step of the game, and each time-step contains information about the events at that time-step, Mario's x-position at that time-step, the player's demographics, and the level being played, making up 37 columns in total.
Columns 0-30 represent in-game events pulled from the raw logs, and are represented by ones and zeros.
We give the full list of all the events in the Appendix.

Columns 31-34 contain player demographic information about their familiarity with games, ranging from 0 (high familiarity) to 4 (low familiarity). This information contains how recently they played \textit{Super Mario Bros.}, any other \textit{Mario} platformer, any platformer, and any video game.
Column 35 contains the level index (ranging from 0 to 15), which we included to help clue the model in on what level was being currently played, even if it was ambiguous from the given level chunk (described below). 

Column 36 contains Mario's x-position at the given time-step. Mario's x-position is not provided by the dataset, but we wanted to use it to provide localized level structure information based on his location, so instead we calculate it by using the log information. We track Mario's predicted position by increasing or decreasing our estimate based on Mario's movement to the right or the left, and the speed of that movement.

Lastly, we crop the logs each play session to the length of the shortest play session (904 time-steps per level) based on prior work \cite{liao2017}. The size for the log array representing each level played is 37x904.

\subsection{Data Parsing - Level Chunks}
\cite{guzdial2016game} provides Java-representations of all 16 levels used in the study. We represent the level structure information with a matrix where each in-game ($x$,$y$) coordinate corresponds to a position in a matrix. 
Each ($x$,$y$) coordinate contains a one-hot array signifying the block present at that position, where each entry corresponds to a block type, with 17 block types in total, given by \textit{Infinite Mario}. We use a one-hot encoding, the array values are all ``zero", except for exactly one ``one" entry in the position associated with the block present at that ($x$,$y$) coordinate. 

We extract the level structure information from the provided files to create the array of all levels. We cut the length of all of the levels to the length of the shortest in the set (198 tiles) based on prior work \cite{liao2017}. To help the model generalize and to conserve computing resources, we remove the bottom row which is identical to the second-bottom row, as well as the top three rows, which are extremely sparse.
The final dimensions of each level's array representation is 198x10x17. 

Unlike prior work, we did not plan to feed in entire levels \cite{liao2017}. 
Rather we wanted to have our model focus on what the player was seeing at a particular moment of gameplay. 
Each ``frame" or ``chunk" data point consists of 10x10x17 slices from our level array, each representing a 10x10 chunk of what is present in a small portion of the level surrounding Mario's calculated x-position.

\subsection{Model Input}
%
Our CNN analyses sequences of ten time-steps at a time, so for each time-step we collect the nine proceeding time-steps worth of logs and append them together. This leaves us with ten (time-steps) by 37 (events) arrays for each data point.

We also collect localized level structure for level analysis based on Mario's predicted x-position. A 10x10x17 ``chunk" (a snapshot of level structure information) is a slice of the level array described in the above Data Parsing - Level Chunks subsection. This is done for the first, middle, and final time-step in our 10-time-step range to capture the map structure as Mario moves through it. This leaves us with a (3x10x10x17) array for each data point. This step in particular differentiates our work from other approaches, particularly from \cite{liao2017}.
We feed both of these inputs, the (10x37) log array and the (3x10x10x17) level chunks matrix into our model's two separate heads. 


\subsection{Model Architecture}
Our CNN is trained on the logs and level data described above. 
Figure \ref{fig:model_visualization}
provides a visualization of the architecture of our model. 
Note that the level visualizations in the figure are a visual approximation of the level data included for visual clarity, and our model does not interpret screenshots of gameplay. 
Our convolution layers all use rectified linear unit (ReLU) activation functions, which we chose because it works well with sparse data and it is computationally efficient.
We used tflearn, a wrapper for TensorFlow, and so had to input our data in a single, combined representation. 
However, we immediately split the input into its two constituent parts and feed it through the two heads described below.

\subsection{Model Architecture - Logs Head}
We input our log data as a 37x10 array into the logs head.
This is passed through a convolution layer with eight 5x5 filters, followed by a 3x3 max-pool layer, and then again through a convolution layer with sixteen 3x3 filters. It is worth mentioning that while using a 3x3 max-pool on a 37x10 layer, we would expect the resulting layer's size to be 12x3, but due to a quirk with tflearn's MaxPool1D function that we used, it further reduces the size to 12x2. The resulting values are flattened, and fed back into the final layers. We chose this architecture as it condenses the information of a number of time-steps and events together, allowing our relatively sparse event logs to be more concisely represented with more generality. This allows more general patterns of log data to be learned, reducing the likelihood of overfitting. 

\subsection{Model Architecture - Level Chunks}
Our level information was input as three 10x10x17 chunks. 
These chunks are fed through a 5x5 convolution layer with 8 filters, then a 2x2 max-pooling layer, and then a 5x5 convolution layer with 8 filters. The resulting output flattens the output into 600x1 (3 chunks worth of the 5x5x8 output) and feeds it into the final layers. We chose this architecture for the same reasons outlined in the previous subsection.

\subsection{Model Architecture - Final Layers}

The separate logs and chunks outputs are concatenated back together and fed through a dropout layer with a 98\% keep rate. This was meant to prevent our model from overfitting, and improves its generality as a result. This is then passed into a fully-connected layer of size 632, with ReLU activation (for convergence efficiency), and then into a final 3 outputs, using softmax activation. The three outputted values represent predictions of player rankings for the levels, which match the three dimensional data from the training set we use provided by \cite{guzdial2016game}. The original dataset had a number of experiential features (fun, frustrating, etc.) and we train models on each of these separately, as in \cite{liao2017}.

Our network is trained using the Adam optimizer, with a batch size of 32 and a learning rate of 7e-5 over 15 epochs. We chose these hyperparameters as we found experimentally that they allowed our model to converge reasonably quickly, and that using smaller learning rates or more epochs did not increase the final accuracy. Training our model takes 45 minutes on a single machine using a 3.5GHz Intel i7-4770K CPU, taking a few hours total to train one model for each experiential features separately.

%
%
%
%


\section{Experiments}
To evaluate our system, we apply it to a heldout test set of the  \textit{Super Mario Bros.}-inspired \textit{Infinite Mario} data \cite{guzdial2016game}, the \textit{Super Mario Bros.: The Lost Levels}-sinpired \textit{Gwario} data \cite{siu2017evaluating}, and the levels from the original \textit{Super Mario Bros.} game. 
Notably, we train our model only on the \textit{Infinite Mario} data, and investigate how well it generalizes to the other, similar game levels. Further, we test our system for these other domains without the use of log information, which is a unique contribution of our paper. 
For each experiment we outline the experimental setup, and then present and interpret the results in the same subsection.



\subsection{\textit{Infinite Mario Bros.} - Experiment}
Our first evaluation is performed on a clone of \textit{Super Mario Bros.} called \textit{Infinite Mario}. We use the dataset provided in \cite{guzdial2016game} which we describe below. We employed an 80-10-10 train-val-test split while training our network.
The dataset includes level structure information for Level 1-1 from \textit{Super Mario Bros.}, as well as 15 procedurally generated levels. It also includes gameplay logs of 75 human players, who first played Level 1-1, and then played two other levels assigned at random. The players were then asked to rank the three levels they played, based on fun, frustration, challenge, design, and creativity. Players ranked the levels by preference, as ``most fun", ``middle fun", and ``least fun" (or replace fun with challenging, frustrating, etc.,) resulting in each ranking being assigned exactly once per player.

As described in our System Overview section, we compiled the level logs for all play sessions (cropped to 904 time-steps per level) into an array, and used 10-time-step sequences, paired with local level structure data from the start, middle, and end time-steps of each sequence as our data points. Notably, since the play sessions are all in a single array, there are 10 ``crossover" data points per level play session, which contain data from the proceeding level play session. We consider this to be a form of data augmentation, as it allows us to use the final 10 time-steps from each play session that might otherwise be discarded. We end up with an even split for the number of data points for each ranking, and over 200,000 data points in total. The goal of our model was, given a single data point, to predict the ranking that the player of that data point assigned to the played level. Each player played three levels, and ranked the levels in order from ``most fun, mid fun, least fun" (and frustrating / challenging respectively) based on how much the level evoked that emotion. Given the even distribution of rankings, we assume that a pure random system would perform with $33.\overline{3}$\% accuracy.

\subsection{\textit{Infinite Mario Bros.} - Results}
\begin{table}[]
    \centering
    \begin{tabular}{c | c | c | c}
        \hline
          Model & Fun & Frustration & Challenge \\
         \hline
         Logs + Level & $91.67$\% & $90.14$\% & $92.86$\% \\
         \hline
         Level Only & $47.32$\% & $51.51$\% & $53.69$\% \\
         \hline
    \end{tabular}
    \caption{\small Test set accuracy for the prediction of player level rankings, based on a sequence of 10 time-steps of game logs and three related level chunks. Performed on \textit{Infinite Mario} play data from \cite{guzdial2016deep}. We include results from our model trained using log and level structure information, as well as a version of our model which is only trained using level structure information as a baseline comparison. \normalsize}
    \label{tab:ResultsTable}
\end{table}

We report our model's test-set accuracy for predicting player self-reports of fun, challenge, and frustration, based on 10 time-steps worth of game log data and three related level chunks in Table \ref{tab:ResultsTable}. Our model is able to correctly predict player level rankings with over 90\% test accuracy for the three metrics we measure. This is a promising improvement over the state of the art on this dataset, achieved by \cite{liao2017}, who use a similar system architecture, and achieve a prediction accuracy of 81.8\%. Their work shows that the use of logs in concert with level structure information improves prediction accuracy, and our results further suggest that using localized level structure data can further improve the results. We also include results of our model trained using only level structure information, in order to highlight the value of training with both logs and level structure information. The bottom row of Table \ref{tab:ResultsTable} shows lower prediction accuracy if the model is trained on only level structure information.

\begin{figure}
    \centering
    \includegraphics[width=\columnwidth, keepaspectratio]{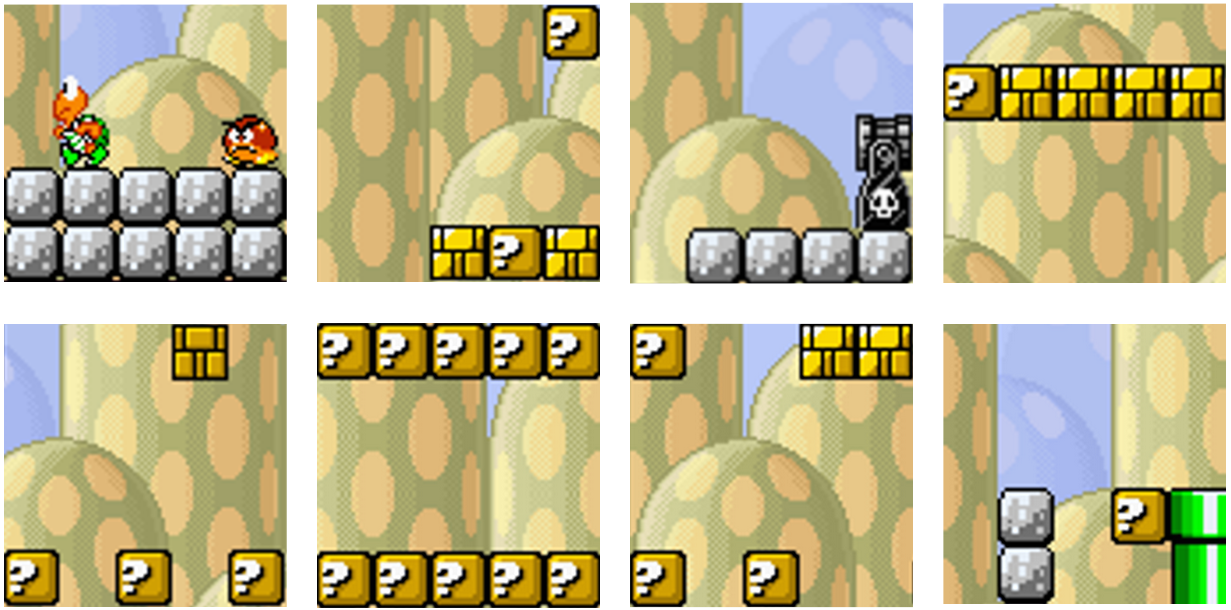}
    \caption{\small Visualizations of maximally-activating 5x5 chunks found in the \cite{guzdial2016game} levels. Each of these chunks maximally-activates one of the eight filters for predicting fun on a level. These filters are from the first convolution layer for the level chunk portion of our CNN. \normalsize}
    \label{fig:MaxFilters}
\end{figure}

A useful feature of our architecture is that it can identify specific level structure patterns which are predictive of specific rankings, and therefore potentially useful to designers. For example, we collected the set of all maximally activating 5x5 chunks for each filter from each level for predicting fun, resulting in a set of 128 chunks. In Figure \ref{fig:MaxFilters} we visualize 8 of those chunks. These indicate level structures which may be valuable for level designers to consider when designing future levels in terms of fun.

\subsection{\textit{Gwario} - Experiment}
For the second experiment we adapt the dataset provided by \cite{siu2017evaluating} of levels and gameplay logs of \textit{Gwario}, a game based on \textit{Super Mario Bros.: The Lost Levels} to test our system. \textit{Gwario} has some changes which alter the gameplay from the game it is based on, but the level structure elements remain largely the same.

The dataset includes results from a human study where players were asked to play two levels from a pool of four (altered levels from \textit{Super Mario Bros.: The Lost Levels}), and rate them based on fun, challenge, and frustration. They also provide the level structure information for the four levels used in the study. We used the player ratings to approximate rankings, given that our model was trained to predict rankings. The levels were rated on a scale of 1 to 5, so we adapted the ratings to our preferential ranking system by, for each metric, taking the mean rating for each level, and then assigning a single ``most" and ``least" ranking to the levels with the highest and lowest mean rating, and a ``mid" rating to the two other levels for that metric, based on the assumption that levels with higher overall ratings would be placed preferentially above lower rated levels.

The \textit{Gwario} levels contain new blocks and enemies that are not present in our original model, so in adapting the levels to work with our model, we replace new elements with similar ones from our model. We replace paratroopa and red koopa enemies with green koopa enemies, we remove piranha plants, and we replace stair blocks with rocks. Additionally, we crop each level to the length of the shortest level in the set (172 blocks), and the level height to 10 blocks by trimming off the bottom row and the upper three rows.

Given the lack of logs, we creative artificial ``empty'' logs. The logs contain all zeros, with the level tag for each time-step being randomly set from zero to fifteen, and Mario's x increasing by one in each time-step. This results in having 172 data points per level, for a total of 688 data points. We test the robustness of our model, which was originally trained using log data, but is tested with empty log data. 

\subsection{\textit{Gwario} - Results}
%

\begin{table}[h]
    \centering
    \begin{tabular}{c | c | c | c | c}
        \hline
          Metric & Most & Mid & Least & Total \\
         \hline
         \textbf{Fun}& & & \\
         Accuracy & $26.24$\% & $59.27$\% & $27.71$\% & $38.7$\% \\
         Pred. rate & $38.85$\% & $36.63$\% & $24.52$\% & $100$\% \\
         \hline 
          \textbf{Frustration} & & & \\
         Accuracy & $22.36$\% & $47.28$\% & $42.11$\% & $41.21$\% \\
         Pred. rate & $23.78$\% & $73.41$\% & $2.81$\% & $100$\% \\
         \hline 
         \textbf{Challenge} & & & \\
         Accuracy & $45.63$\% & $48.99$\% & $27.39$\% & $36.48$\% \\
         Pred. rate & $15.21$\% & $29.25$\% & $55.54$\% & $100$\% \\
         \hline 
         
    \end{tabular}
    \caption{\small Predictions for individual metrics on the \textit{Gwario} dataset, predicted by our model trained with logs. Includes the frequency at which each rating was assigned, and the accuracy of those ratings. Among the 688 data points, one quarter were ranked most, half were ranked mid, and one quarter were ranked least for each metric.\normalsize}
    \label{tab:GwarioTable}
\end{table}

\begin{table}[h]
    \centering
    \begin{tabular}{c | c | c | c | c}
        \hline
          Metric & Most & Mid & Least & Total \\
         \hline
         \textbf{Fun}& & & \\
         Accuracy & $33.33$\% & $49.66$\% & $28.75$\% & $46.97$\% \\
         Pred. rate & $1.33$\% & $86.85$\% & $11.82$\% & $100$\% \\
         \hline 
          \textbf{Frustration} & & & \\
         Accuracy & $25.73$\% & $55.77$\% & $18.18$\% & $27.92$\% \\
         Pred. rate & $90.69$\% & $7.68$\% & $1.62$\% & $100$\% \\
         \hline 
         \textbf{Challenge} & & & \\
         Accuracy & $25.55$\% & $50.48$\% & $14.29$\% & $29.39$\% \\
         Pred. rate & $83.46$\% & $15.51$\% & $1.03$\% & $100$\% \\
         \hline 
         
    \end{tabular}
    \caption{\small Predictions for individual metrics on the \textit{Gwario} dataset, predicted by our model trained without logs. This includes the frequency at which each rating was assigned, and the accuracy of those ratings. Among the 688 data points, one quarter were ranked most, half were ranked mid, and one quarter were ranked least for each metric.\normalsize}
    \label{tab:GwarioTable2}
\end{table}

Our evaluation of the Gwario dataset for predicting player rankings is shown in Table \ref{tab:GwarioTable}. We achieve an overall prediction accuracies of $38.7$\%, $41.21$\%, and $36.48$\%, for Fun, Frustration, and Challenge respectively, which are all above an expected random baseline performance. This suggests that our model was able to learn some patterns associated with the measured metrics. 

Regardless, these values could suggest an issue with a number of the assumptions we made, including the assumption that a 1-5 scoring system rating can be converted to preferential ranking, that ranking these levels with one ``most fun", two ``mid fun", and one ``least fun" might not be appropriate, or that the overall rating of a level can be generalized across the entire span of a play session.

Table \ref{tab:GwarioTable2} is similar to Table \ref{tab:GwarioTable}, but contains predictions using our model architecture trained on only level structure information, without logs. We provide this as an additional baseline for comparison to our model trained on both logs and level structure information. The model trained without logs achieves prediction accuracy values of $46.97$\%, $27.92$\%, and $36.48$\%, for Fun, Frustration, and Challenge respectively. This baseline thus outperforms our proposed model at predicting Fun, which we anticipate is because the baseline predominantly predicts a single rating for each metric. This could indicate that the baseline is identifying that the Gwario levels are very similar to one another, which could suggest that a level parser trained without log data can provide more accurate results in a different domain than the one it is trained on.



\subsection{\textit{Super Mario Bros.} - Experiment}
Finally, we test the ability of our model to learn meaningful features by testing it on a number of levels from the original \textit{Super Mario Bros.} We use levels provided by \cite{VGLC} in the Video Game Level Corpus (VGLC), which contains level structure information for 15 levels from the original \textit{Super Mario Bros.} game. We order the levels in the order that they appear in game, and assign them level numbers based on their placement in this order, with a level's level number increasing the later in the game that it appears.

The original \textit{Super Mario Bros.} does not differentiate between unbreakable block types, so we convert unbreakable blocks along the bottom of the screen to ``hilltop" blocks (which appear  along the bottom of the \textit{Infinite Mario} training set levels) and all other unbreakable blocks into ``rock" blocks. We also crop each level to the length of the shortest level in the set (150 blocks) and the height of each level to 10 blocks by removing the top four rows.

We use an identical ``empty" logs system as described above in the Gwario - Experiment subsection to create blank gameplay logs. Because the levels do not have player ratings, we instead assume that levels from \textit{Super Mario Bros.} would appear in order of increasing difficulty, and test whether our model will predict increasing challenge rankings for each level in succession. We record the mean challenge prediction across all data points for each level, and use a Spearman's rho correlation test with these values to see if level number and predicted challenge are correlated.

\subsection{\textit{Super Mario Bros.} - Results}
\begin{table}[]
    \centering
    \begin{tabular}{c | c | c | c}
        \hline
          & Most & Mid & Least \\
         \hline
         $r_s$ & $0.8143$ & $0.75$ & $-0.7607$ \\
         \hline
         $p$ & $2.194e{-4}$ & $1.281e{-3}$ & $9.911e{-4}$ \\
         \hline
         $95\%$ CI & $[0.57, 0.94]$ & $[0.39, 0.91]$ & $[-0.92, -0.41]$ \\
         \hline
    \end{tabular}
    \caption{\small Spearman's rho correlation test results measuring correlation between predicted challenge and a level's order of appearance in \textit{Super Mario Bros.} Positive (or negative) r$_s$ values indicate those ratings become more (or less) common the later in the game that the rated level appears in, and greater r$_s$ magnitudes signify stronger correlation. 95\% confidence intervals are also included. \normalsize}
    \label{tab:MarioStatistical}
\end{table}

We present the results from a Spearman's rho correlation test for determining a relationship between level number and challenge in Table \ref{tab:MarioStatistical}. The correlation for ``most" and ``mid" are positively correlated with level number, and the correlation for ``least" is negatively correlated with level number, signifying that ``most" and ``mid" rankings increase alongside level number, while ``least" rankings decrease.

Notably, all three correlations have a p-value of $<0.01$, signifying that the correlation between our model's prediction of difficulty and the order that the levels appear in the original \textit{Super Mario Bros.} is statistically significant.  If our assumption holds that the difficulty of levels in \textit{Super Mario Bros.} increases throughout the game, then that means that our model was able to learn to detect useful features for detecting challenge in Mario levels, even without access to gameplay logs on those levels, and despite having never seen these levels before (other than Level 1-1). 
To the best of our knowledge, we are the first to show the ability of a model trained on \textit{Infinite Mario Bros.} log and level data can demonstrate significant predictive performance on the original game. 



\section{Limitations}
There are a number of limitations with the approaches that we took, which we believe offer avenues for future work in this domain.

The way that we calculate Mario's x-position from the level logs does not account for acceleration (and deceleration), and instead tracks all movement as movement at maximum speed. Additionally, it does not detect if Mario's movement would be blocked, making his actual position and his predicted position mismatch when that happens. Creating a system which used the level structure or even an A* planning agent could improve the accuracy of our approximated Mario position, but recording this information at playtime would ultimately provide the most accurate data.

Another issue with our work is the assumption that player experience ratings for an entire level are representative of their experience across their entire play experience. We hope that the results of our \textit{Super Mario Bros.} experiment show that despite this, it seems that some meaningful features are being learned by our model.

We compare the performance of our model to a random baseline due to time constraints. Comparing to the performance of a different model would better demonstrate the relative performance of our approach.

We use a cross-validation system using a 80-10-10 split, rather than a k-fold test, due to time and computing resource constraints. Using a k-fold test could better demonstrate that our model's presented performance is representative of it's actual ability. 


We hand-edit some level elements from the \textit{Gwario} and \textit{Super Mario Bros.} datasets to replace elements that are not present in the test data with similar elements that are. This could be avoided by instead including an ``unknown" block symbol for blocks that do not appear in the dataset.


The visualizations that we presented for chunks which maximally activated filters for fun prediction were hand-selected from a larger set. The chunks in this set are still valuable, as they highlight specific chunks for human appraisal, reducing designer burden.
In the future, we hope to use these identified chunks or the model itself as part of a level generation pipeline.


\section{Conclusions}
In this paper we presented a means of predicting player affect through the use of gameplay logs and localized level structure information, which outperforms past work. We ran several experiments to test the effectiveness of our system, using datasets other than the one it was trained on, and we provided evidence that our system is able to learn to identify meaningful level structure patterns. Our system leveraged local chunks of levels and gameplay logs to learn level features, but does not require gameplay logs after training. 
To the best of our knowledge, this is a novel approach, and one that could help improve cross-domain player modelling.


\bibliography{main}
\newpage
\appendix

\section{\textit{Infinite Mario} Events}

\textit{Infinite Mario} events listed in order are: StartLevel, WonLevel, LostLevel, Jumping, RightMove, LeftMove, Running, Ducking, Little, Large, Fire, DieByGoomba, DeathByShell, DeathByBulletBill, DieByGreenKoopa, DeathByGap, UnleashShell, BlockCoinDestroy, BlockPowerDestroy, FireKillGoomba, StompKillGoomba, StompKillGreenKoopa, ShellKillGoomba, ShellKillGreenKoopa, FireKillGreenKoopa, CollectCoin, BlockPowerDestroyBulletBill, StompKillBulletBill, ShellKillBulletBill, BlockCoinDestroyBulletBill, and CollectCoinBulletBill. Events 4-10 (RightMove through Fire) are continuous events, and while they are recorded in the logs only when these events begin and end, we instead record Mario's state in all time-steps. So if Mario becomes Large Mario, we set the Large event to active until Mario changes to Small Mario or Fire Mario.

\begin{table}[h]
    \centering
    \begin{tabular}{c | c | c | c | c}
        \hline
        Order & Level Name & Most & Mid & Least \\
        \hline
         0 &   1-1 &   2.66\%  &  4.37\%  &  92.96\% \\
         1 &   1-2 &   14.75\% &  9.59\%  &  75.64\% \\
         2 &   1-3 &   32.91\% &  17.41\% &  49.66\% \\
         3 &   2-1 &   39.51\% &  60.48\% &  0.00\% \\
         4 &   3-1 &   44.24\% &  24.92\% &  30.82\% \\
         5 &   3-3 &   51.74\% &  24.57\% &  23.67\% \\
         6 &   4-1 &   55.67\% &  26.60\% &  17.72\% \\
         7 &   4-2 &   57.89\% &  28.52\% &  13.58\% \\
         8 &   5-1 &   60.24\% &  29.58\% &  10.16\% \\
         9 &   5-3 &   63.14\% &  29.41\% &  7.43\% \\
         10 &  6-1 &   64.19\% &  29.43\% &  6.36\% \\
         11 &  6-2 &   61.30\% &  34.08\% &  4.60\% \\
         12 &  6-3 &   60.62\% &  35.41\% &  3.95\% \\
         13 &  7-1 &   59.81\% &  36.74\% &  3.43\% \\
         14 &  8-1 &   57.37\% &  38.95\% &  3.66\% \\
    \end{tabular}
    \caption{Predicted challenge rankings for levels from the original \textit{Super Mario Bros.}. Levels are numbered in increasing order based on their order in-game. Interestingly, level 3 in the set (level 2-1 in \textit{Super Mario Bros.}) seems to be an outlier, having no ``least challenging" ratings, and a large jump in higher challenge ratings. This may be due a shortcoming in our model, but could also indicate that Mario 2-1 is much more difficult than other levels nearby in the ordering. }
    \label{tab:RawMarioResults}
\end{table}

\end{document}